%% file: CCIS_draft.tex
\documentclass[runningheads]{llncs}
\usepackage[T1]{fontenc}
\usepackage{graphicx}
\usepackage{amssymb}
\usepackage{hyperref}
 \usepackage{tcolorbox} 
\usepackage{mdframed}  
\usepackage{array}

\usepackage{color}

\newcolumntype{P}[1]{>{\centering\arraybackslash}p{#1}}

\newtcolorbox{takeawaybox}[1][]{
  colframe=gray!75,      
  colback=white!10,       
  coltitle=black,        
  boxrule=0.5mm,         
  fonttitle=\bfseries,   
  title=#1               
}

\input{tables/results}

\input{tables/methodology}
\input{tables/background}

\begin{document}
%
\title{VAPU: System for Autonomous Legacy Code Modernization}
%
%
\author{Valtteri Ala-Salmi\inst{1}
\and
 Zeeshan Rasheed\inst{1}
 \and
 Abdul Malik Sami\inst{1}
\and
Muhammad Waseem\inst{1}
\and
 Kai-Kristian Kemell\inst{1}
 \and
 Jussi Rasku\inst{1}
 \and
 Mika Saari\inst{1}
 \and
 Pekka Abrahamsson\inst{1}
}
\authorrunning{V. Ala-Salmi et al.}

\institute{Faculty of Information Technology and Communication Science, Tampere University, Finland 
\email{\{valtteri.ala-salmi, zeeshan.rasheed, malik.sami,muhammad.waseem, kai-kristian.kemell, jussi.rasku, mika.saari, pekka.abrahamsson\}@tuni.fi}\\}
\maketitle              
\begin{abstract}

 LLM-based multi-agent systems have become an emerging technology to enhance LLM's capabilities in complex multi-phased tasks. One recently studied area of these systems has been code generation, which includes modernization of legacy applications. Applications, especially on the Internet, often have deprecated components related to compatibility, security, and reliability that cause risks in their usage. Due to the high costs in resources, companies hesitate with the update, as such, a system capable of assisting in a low cost and reliable way is required. To this end, we take a step forward to integrate an LLM-based multi-agent system as part of an legacy web application update to provide a cost-effective solution to update legacy applications autonomously. We propose a multi-agent system named a Verifying Agent Pipeline Updater (VAPU), which is designed to update code files in phases while simulating different roles in a software development team. 
 
 In our previous study, we evaluated the system for legacy version updates by using six legacy web application view files by resulting errors and accomplished requirements. This study extends the previous evaluation of a multi-agent pipeline system by extending the evaluation of VAPU from a single LLM to five LLMs and using the temperature parameter in both 0 to 1 settings. Additionally, we tested the system with 20 open-source Python GitHub projects. The results of the evaluation were compared to Zero-Shot Learning (ZSL) and One-Shot Learning (OSL) prompts. The extended evaluation of VAPU showed that particularly in a low-temperature VAPU can get similar level of error count compared to the ZSL/OSL prompts but with a higher level of fulfilled requirements, depending on the LLM. VAPU showed up to 22.5\% increase in the succeeding Python file update requirements compared to ZSL/OSL prompts. The study indicates that an LLM-based multi-agent system is a capable solution to update components of a legacy application autonomously. This allows practitioners to improve the cost-effectiveness of legacy applications' maintenance, and making it a more compelling choice. VAPU provides a functional foundation for practitioners to produce more advanced systems to update legacy applications. The code is publicly available on GitHub \url{https://github.com/GPT-Laboratory/Software-Refactoring}.
 
\keywords{Generative AI \and Large Language Models \and Legacy Systems \and Multi-Agent System \and Prompt Engineering \and Temperature}
\end{abstract}
\section{Introduction}
\label{sec:introduction}

The evaluation of LLM-based multi-agent systems for solving multi-phased processes in code generation has become an emerging topic \cite{fourney2024magentic,huang_agentcoder_2024,lei2024planning}. This has been possible due to the transformer architecture in LLMs, which was introduced by Vaswani \textit{et al}. \cite{vaswani2017attention}. The transformer architecture has allowed LLM's size to be efficiently scalable due to its self-attention mechanism, which allows parallelized computation in graphical/tensor processing units \cite{vaswani2017attention}. As LLM performance is connected to its parameter size \cite{kaplan_scaling_2020}, larger and better performing LLMs give new possibilities for multi-agent systems with more complex tasks.

One notable challenge in software engineering is the amount of legacy applications in business. According to Demir \textit{et al}. \cite{demir2021our}, 95\% of the 5.6 million analyzed web applications contain a deprecated component. Keeping applications up-to-date takes an estimated 40\%-90\% of a company's resources \cite{davis_97_2009,erlikh_leveraging_2000}. As such, companies can easily accumulate technical debt for business critical software applications which then can become legacy systems. Updating legacy systems is a difficult process for companies due to the expensive cost of resources that can increase multiple times over the initial budget \cite{galinium2012success}. The challenges in safety \cite{smyth_penetration_2023}, usability \cite{ali_addressing_2020}, and compatibility \cite{antal2016transforming} are expected because companies are reluctant to risk a cost-ineffective operation and choose to conduct business with outdated applications.

To address these challenges in legacy system updates, automated code generation using multi-agent systems presents a promising solution due to a high coding success rate compared to other LLM-based methods \cite{huang_agentcoder_2024} and manageable cost \cite{rasheed2024codepori}. This work extends the findings of Ala-Salmi \textit{et al}. \cite{enase25}. We conducted a more comprehensive evaluation of a multi-agent pipeline, now properly named as Verifying Agent Pipeline Updater (VAPU), to provide additional information on its capabilities. We proposed VAPU to provide a cost-effective solution to update legacy application files. VAPU conducts a complex set of operations in phases by using multiple agents, each with a specific task in the process. For example, the verification agent gives feedback after every accomplished phase to evaluate its success, and the finalizer agent will then finish the phase based on that evaluation. 

VAPU's functionality was first verified with six view files from a legacy web application that needed an update in it's web framework version \cite{norri_digitization_2020}. The amount of LLMs used was extended from one to five, including Claude 3.5 Sonnet, DeepSeek-V3, GPT-4o mini, GPT-4o, and Nova Pro 1.0. The tests were also repeated at zero-temperature for less randomness. VAPU was then additionally validated with 20 Python open-source GitHub projects. The results showed that the medium parameter models Nova Pro 1.0 and DeepSeek-V3 excelled with low error rate, while higher parameter models Claude 3.5 Sonnet and GPT-4o excelled in the correct outcomes in harder tasks. At zero-temperature, VAPU had a similar error rate compared to OSL/ZSL prompts while excelling in Python tasks up to 22.6\% depending on the LLM.

The results indicate that a multi-agent system is capable of updating legacy application files and provides a considerable solution for a cost-effective update of an application. VAPU provides a framework for practitioners to further refine a system suited to update legacy files. In addition, data of VAPU's behavior in different parameters provides insight into multi-agent systems behavior compared to one inference call alternatives, such as OSL/ZSL prompts for academic study.

The \textbf{contributions} of this study can be summarized as follows:

\begin{itemize}
    \item Extending the multi-agent system's evaluation from one LLM model to five to get information of the system's behavior and performance differences across different models. 

    \item Comprehensive evaluation of the multi-agent system across different configurations and datasets, including zero-temperature testing to reduce randomness and validation with 20 Python open-source projects for broader applicability assessment.

    \item We demonstrate that an LLM-based multi-agent system can update code at a similar level of error rates as OSL/ZSL prompts, with a higher possibility of a successful update depending on the used LLM.

    \item We publicly released the multi-agent system used and the evaluation results dataset to access all collected data to validate our study \cite{VAPU,tampere_university_2025_16790515}. 
    
\end{itemize}

The rest of the paper is organized as follows: Section \ref{Background Study} presents a background study on code generating multi-agent systems and how such a system could be used to solve challenges in updating legacy applications. Section \ref{Research Method} explains the methodology of this paper and presents the multi-agent system used. This is followed by Section \ref{Results} which presents the results obtained in this study. Section \ref{Discussion} discusses the results obtained from the study. Section \ref{Conclusion} concludes this study.

\section{Related Work}
\label{Background Study}

This section presents previous studies using LLMs to generate code. Section \ref{section:codegeneration} discusses previous LLM-based systems created for code generation. Section \ref{section:LLMchallenges} discusses the challenges found in code generation using LLMs. Section \ref{section:challengeslegacy} discusses the challenges in modernizing legacy systems.

\subsection{Code Generation Using LLM-Based Systems}
\label{section:codegeneration}

This section introduces four LLM-based code generating systems where the studies have used a coding benchmark HumanEval \cite{chen2021evaluating} to generate Python code for different problems using GPT LLMs. We then compare them with baseline values found in the studies and the literature to validate their value in code generation.

Reflection system proposed by Shinn \textit{ et al}. \cite{shinn2024reflexion}, has an evaluator component that gives a numerical evaluation of the agent output in the given operation. This is then processed in a self-reflection system, which added textual reflective analysis in the agents long memory to be used in the next outputs. Huang \textit{et al}. \cite{huang_agentcoder_2024}, proposed a system called AgentCoder in which one agent is tasked with producing code, the second one is tasked with inventing test cases, and the last agent
executes the test cases and provides results. Zhong \textit{et al}. \cite{zhong_debug_2024}, proposed a system named Large Language Debugger (LDB), in which an agent-generated application is divided into control blocks where individual test cases are created and executed for each block. The LLM-based debugger then receives the intermediate values of the blocks and uses them to determine the correctness of the code \cite{zhong_debug_2024}. Lei \textit{et al}. \cite{lei2024planning} proposed a system named large language model programming workflow (LPW), which is a two-phase pipeline system with the first phase generating the code and the second phase testing and refining the code.

The comparison of LLMs using the systems and the baseline values is visible in table \ref{tab:humaneval}. Based on the results of the literature, the studied systems showed improvement to the baseline in each GPT-model and in each system. As such, the literature supports the benefits of using a LLM-based system for code generation including multi-agent systems \cite{huang_agentcoder_2024}. It is important to note that the HumanEval result can vary highly between the test cases, as can be seen from the table. The variation might be explained by the frequent update of the GPT-models \cite{openai_Model_2025} and the stochastic nature of LLMs, which causes a variation in the results \cite{brown_language_2020}.

\humanEvalmetrics

\subsection{Challenges in LLM-Generated Code}
\label{section:LLMchallenges}

This section discusses challenges in code generation using LLMs found in the literature. The following challenges were found in three analyzed studies that focused on recognizing them:

\begin{takeawaybox}[Challenges in code generation using LLM]
\scriptsize
\begin{itemize}
  \item[$\bullet$] The quality of the produced code decreases with the length of the code and the difficulty of the task \cite{chong_artificial-intelligence_2024,dou_whats_2024,liu2024refining}. 
  \item[$\bullet$] LLM does not always consider requirements of the user, or those that are needed for a reliable and safe code \cite{chong_artificial-intelligence_2024,dou_whats_2024,liu2024refining}.
  \item[$\bullet$] LLM can generate more errors in the code when attempting to fix it \cite{chong_artificial-intelligence_2024,liu2024refining}.
\end{itemize}
\end{takeawaybox}

In all studies, a decrease in the quality of the LLM-generated code was found by increasing length or difficulty \cite{chong_artificial-intelligence_2024,dou_whats_2024,liu2024refining}. Liu \textit{et al}. \cite{liu2024refining} found that the probability of the returned code working correctly decreases with harder code tasks and the length of the produced code in different Python and Java programming tasks. Dou \textit{et al}. \cite{dou_whats_2024} solved different coding benchmarks with different LLMs and evaluated results by syntax, runtime, and functionality errors. The study found that the failure rate of tasks increases with the number of code lines, code complexity, and API calls required. Chong \textit{et al}. \cite{chong_artificial-intelligence_2024} found that creating a memory buffer correctly with an LLM had a lower success rate with complex tasks, for example, complex multiplication of two floats, succeeded in 1.5\% cases, while subtracting a float from an integer had a higher success rate of 50.1\%.

The problems of following an user request and producing it as reliable and safe were also found in all studies \cite{chong_artificial-intelligence_2024,dou_whats_2024,liu2024refining}. In Liu \textit{et al}. \cite{liu2024refining} errors such as redundant modifier, ambiguously named variables, and excess local variables were found, which lowers the reliability and safety of the code. Dou \textit{et al}. \cite{dou_whats_2024} found that the most notable reason for the errors were LLM's performance challenges, particularly misunderstanding of the problem provided by the user and logical errors in the code including problems in safety and reliability. Chong \textit{et al}. \cite{chong_artificial-intelligence_2024} found that while LLM generates fewer lines of code,
the code lacks the defensive programming that exists in human-written code, leaving LLM-generated code vulnerable to security threats.

The third challenge found is that while an LLM feedback loop can improve code generation, it
can also hallucinate additional errors in the code \cite{chong_artificial-intelligence_2024,liu2024refining}. Liu \textit{et al}. \cite{liu2024refining} found that giving feedback
improves the LLM-generated code up to 60\% but with the possibility of additional errors
added to the code in both providing textual feedback and slightly less with static and runtime information. Chong \textit{et al}. \cite{chong_artificial-intelligence_2024} found that the LLM can generate new
security problems in the code through a feedback loop and not just remove them,
especially if the file does not contain problems in the first place.

From the literature, a multi-agent system could provide solutions to these challenges.
For the challenge related to increasing code length and complexity, Rasheed \textit{et al}. \cite{rasheed2024codepori} suggested generating code in modules, which allows the code to be separated into smaller units for the agents to generate. Fourney \textit{et al}. \cite{fourney2024magentic} proposed an agent called orchestrator, which creates a highly detailed plan for a provided coding problem, to improve the results in complex coding tasks. With unreliable and unsafe code, Nunez \textit{et al}. \cite{nunez2024autosafecoder} proposed an agent designed to find Common Weakness Enumeration (CWE) problems, which reduced the probability of CWE from 49\% to 36\%. Feedback loops were used in the studied multi-agent systems \cite{huang_agentcoder_2024,lei2024planning,shinn2024reflexion,zhong_debug_2024} in Table \ref{tab:humaneval}. Despite the risks of additional errors, the systems had a higher success rate than baseline LLMs, which suggests that it is more beneficial to include it in the system.

\subsection{Challenges in Upgrading Legacy Applications}
\label{section:challengeslegacy}

This section discusses challenges in legacy application updates found in the literature. Legacy systems are outdated implementations by used technologies and programming languages with hardware, software, or other parts of the system
being outdated \cite{sommerville_software_2016}. The strategies for dealing with legacy systems are disposing of the system, keeping the system as such,
and re-engineering or replacing components in the legacy system \cite{sommerville_software_2016}. This study focuses on application side of legacy systems with re-engineering or replacing components with the help of LLM-based multi-agent system. The following three challenges were recognized based on three studies found in the literature:

\begin{takeawaybox}[Challenges in updating legacy applications]
\scriptsize
\begin{itemize}
  \item[$\bullet$] The programmers can have knowledge gaps in either old or new technologies of a legacy application \cite{de_marco_cobol_2018,fritzsch_microservices_2019}. 
  \item[$\bullet$] Identifying the input, output, and code functionality of a legacy system's component is a demanding, time-consuming task in legacy applications \cite{de_marco_cobol_2018,vesic_framework_2023}.
  \item[$\bullet$] Breaking down the update process into smaller parts and successfully managing them is a challenge in legacy applications \cite{de_marco_cobol_2018,fritzsch_microservices_2019}. 
\end{itemize}
\end{takeawaybox}

The first recognized challenge is the knowledge gaps of developers in new or legacy technology in a project \cite{de_marco_cobol_2018,fritzsch_microservices_2019}. In De Marco \textit{et al}. \cite{de_marco_cobol_2018}, a legacy
mainframe application was migrated to Linux servers with changes to the database
and a transition of the program implementation from COBOL to Java. One challenge of the project was that the team knowledgeable of COBOL had challenges understanding Java and the other team similarly with COBOL, which led to challenges especially when planning new features \cite{de_marco_cobol_2018}. In Fritzsch \textit{et al}. \cite{fritzsch_microservices_2019}, 14 legacy systems in different stages of
migration to microservices were analyzed with interviews with companies. Based on the interviews, the lack of expertise in microservice architecture emerged as the most common technical challenge, identified by 8 out of 14 projects \cite{fritzsch_microservices_2019}.

The second recognized challenge is that identifying the input, output and functionality of each legacy application component is a demanding time-taking task in projects \cite{de_marco_cobol_2018,vesic_framework_2023}. According to De Marco \textit{et al}. \cite{de_marco_cobol_2018}, the testing phase of the system's migration caused a year delay for the project. It was found that there were no high-level tests for components that made understanding the input, output, and inner functionalities of an outdated component a time-consuming task \cite{de_marco_cobol_2018}. In Vesić and Laković \cite{vesic_framework_2023}, a framework was presented to evaluate legacy systems. Using the framework to evaluate an existing information system, the system lacked a proper documentation, lack of knowledgeable personnel of how the system operates, and a poor software architecture that would lead to challenges if the system were ever tried to be modernized \cite{vesic_framework_2023}. 

The third recognized challenge is that breaking down a legacy system modernization project into smaller manageable tasks can be a challenging operation \cite{de_marco_cobol_2018,fritzsch_microservices_2019}. The decomposition of applications into microservices was recognized as a technical challenge in 8 of 14 projects, making it the most common challenge alongside the knowledge of microservices \cite{fritzsch_microservices_2019}.
According to De Marco \textit{et al}. \cite{de_marco_cobol_2018}, the project was divided into smaller work packets. The work packets were used to forecast the project duration, however due to challenges in testing, the later work packets lasted longer, causing delay in the project \cite{de_marco_cobol_2018}. As such, the later packets were not as manageable as the first ones, causing an imbalance for the project.

LLM-based multi-agent systems could offer solutions to the recognized challenges. As LLM's can be trained on different tasks and technologies \cite{brown_language_2020}, a multi-agent system could offer agent's specifying in a certain technology or task to solve knowledge caps. For analyzing the input, output, and functionality of components, a multi-agent system could use I/O specifications as investigated in the literature \cite{wen2024grounding}. Breaking down the legacy system project could be solved by an agent that specializes in division of the task into smaller manageable tasks \cite{fourney2024magentic,rasheed2024codepori}.

Compared to the LLM-based systems examined in Section \ref{section:codegeneration}, VAPU is designed to update code instead of generating it. The system needs to take into account both the challenges in code generation in LLMs and the challenges related to updating legacy code. As such, VAPU is designed to provide benefits of multi-agent systems with a design and implementation that can provide them for updating existing code.

\section{Research Method}
\label{Research Method}

This section presents the methodology and the evaluation framework for the proposed system. Section \ref{RQ} provides the formulated Research Question (RQ). The design and implementation of the proposed system and the used LLMs are discussed in Sections \ref{proposed:sysmte}-\ref{evaluated:llms}. The evaluation framework is discussed in Sections \ref{verification}-\ref{validation}.

\subsection{Research Question}\label{RQ}
In this study, the following RQ is proposed:

\begin{tcolorbox}[colback=gray!2!white,colframe=black!75!black]
\textit{\textbf{RQ.} To what extent are multi-agent systems reliable and effective for updating legacy code?
}
\end{tcolorbox}

The aim of this \textbf{RQ} is to test the reliability of a multi-agent system when tasked to autonomously upgrade legacy web application code. This objective focuses on assessing how reliable the proposed system can update and modify outdated functionalities by refactoring the deprecated code autonomously. 

\subsection{Proposed Multi-Agent System: VAPU}
\label{proposed:sysmte}

The proposed system VAPU is a multi-agent system designed to update code in phases. Originally called as multi-agent pipeline \cite{enase25}, it is now clarified as a VAPU to make a distinction from other systems with a pipeline architecture. The proposed multi-agent system uses software development team agents to go through a development process, which has obtained good results in previous multi-agent systems \cite{huang_agentcoder_2024,rasheed2024codepori}. 

The proposed system VAPU is visible in Figure \ref{fig:fig_model}. The first input of the system consists of a list of requirements or a project description. The second input is the original codebase, which includes the code that the user wants to update. The output is the updated codebase, which is modified based on the requirements file. The requirements can specify any request to modify or update the code. For example, the requirement could be to update the code compatibility from version X to version Y or to replace any library that the code uses. The system is
composed of four units, all tasked with a different role in the code updating process. VAPU uses a manager, a verifier, and a finalizer agent in the system with the addition of the developer agents changed into a prompt maker and an execution agent in the component called the task pipeline. The functions of each component are explained below:

\begin{figure*}[!t]
\centering
\includegraphics[scale=0.26, alt={Diagram showing VAPU's process flow. From left, input files are shown, followed by VAPU and relations between agents, ending with the output file on the right.}]{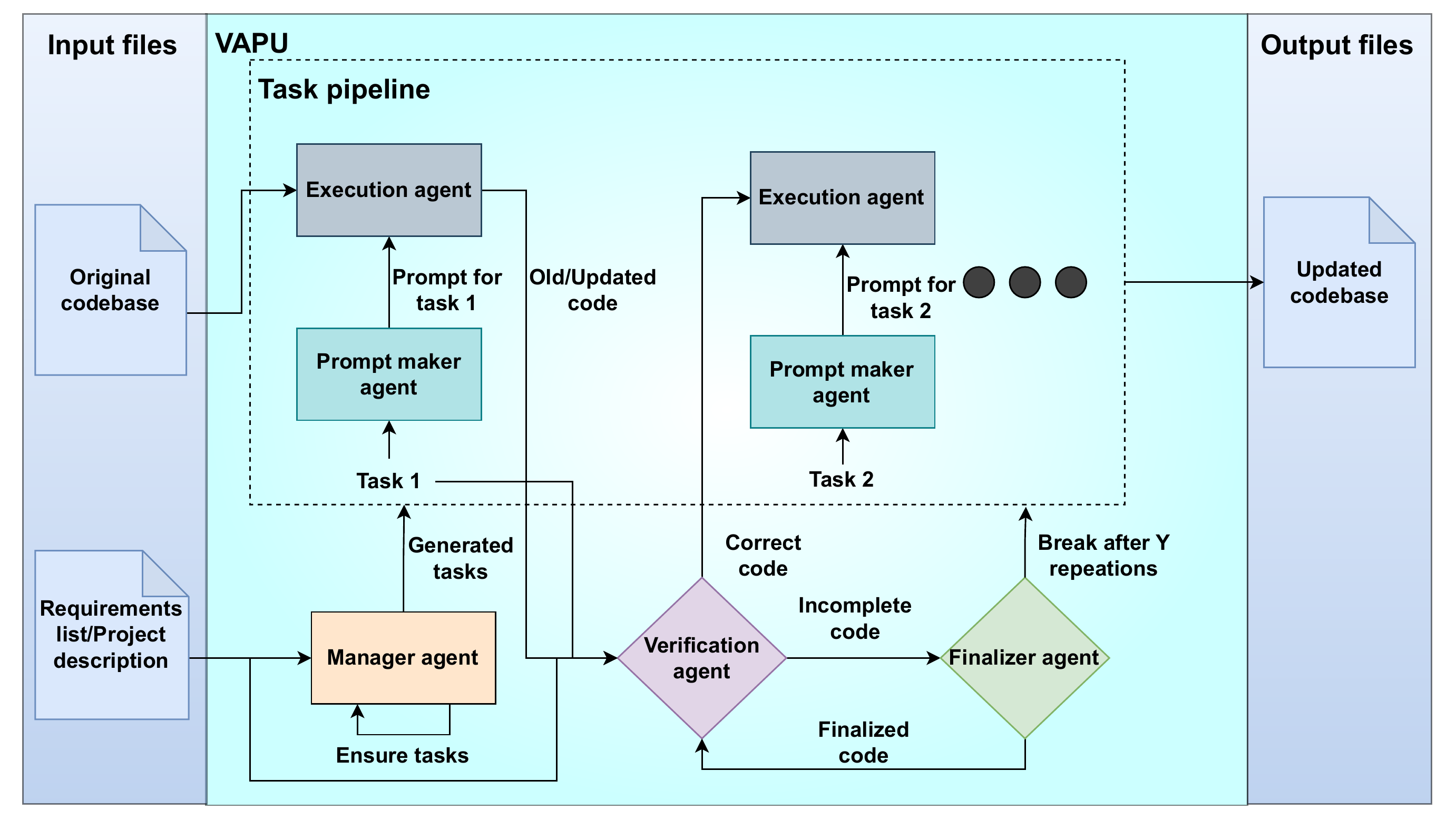}
\caption{Proposed system: VAPU for updating existing code adapted from \cite{enase25}}
\label{fig:fig_model}
\end{figure*}

\textbf{Manager agent:} The manager agent receives the requirements or a project description
from the user and writes them into manageable operations in chronological execution order. The tasks are written in abstract level, which are
later defined properly in instructions by the task pipeline, where the manager agent sends the tasks. The manager
agent reflects once before sending the tasks to ensure that the tasks are in
chronological order and are overall related to the requirements to avoid hallucinations.

\textbf{Task pipeline:} The task pipeline consists of two agent types, a prompt maker and an execution agent. First, the prompt maker agent receives the requirement list/project description and a specific task from it from the manager agent. Then the prompt maker agent creates a detailed instruction's to fulfill that task which is used as a prompt. The second agent in the pipeline, the execution agent, then receives this prompt with the original code and performs the requested task. After each task, the code is sent to the verification agent to review the completion of the task. After every task requested by the manager agent is completed in the pipeline, the updated code is returned as an output for the user.

\textbf{Verification agent:} The verification agent ensures that a task has been completed by comparing the result with the code before the task. To determine success, the verification agent receives both the task and the requirements list/project description and ensures that the end result corresponds to them. If the verification agent accepts the task,
it will return the code to the pipeline. If the code is determined to be incomplete, the verification agent will send the code to the finalizer agent.

\textbf{Finalizer agent:} The finalizer agent makes changes to the updated code if
the verification agent notices that the tasked operation has not been completed
successfully. For that, the verification agent provides a list of changes that are needed to do for the code, before the verifier agent can accept it as completed. After making changes, the finalizer agent returns the modified
code back to the verification agent to verify the code again. The finalizer agent and the verification agent form a feedback loop, which either ends when the verifier agent accepts all the changes or a certain number of loops if the verifier agent fails to accept the code. This might happen, for example, if the verification agent starts to hallucinate
and finds problems that do not exist, as found in the background study \cite{chong_artificial-intelligence_2024,liu2024refining}.

The intended benefits of VAPU in code update compared to using an LLM alone are listed below:

\begin{itemize}

\item[1.] \textit{Self-division:} The division of the project into smaller tasks was a recognized challenge for legacy modernization projects in Section \ref{section:challengeslegacy}. Additionally, generating longer code files was a challenge for LLMs in \ref{section:LLMchallenges}. With the division of tasks, the simultaneous amount of produced code and the complexity of a task can be lowered, increasing the excepted quality of the outcome.

\item[2.] \textit{Self-feedback:} Every separate task will be analyzed in the verifier agent
and then possibly improved by the finalizer agent iteratively. This generates a feedback loop where the updated code is expected to improve. This is backed by the findings of Section \ref{section:codegeneration}, where the code metric did increase with the LLM-based systems using a similar implementation. This allows the code to better align to the user requirements, which was recognized as a challenge in Section \ref{section:LLMchallenges}.

\item[3.] \textit{Self-instructive:} Writing complex prompts to instruct LLM is a challenging task for people, especially if they are not familiar with
prompt engineering. According to Zamfirescu-Pereira \textit{et al}. \cite{zamfirescu-pereira_why_2023}, non expert prompt writers were found to have problems with overgeneralization and human interactive style of writing prompts which resulted in low performance. With a self-instructive system, the challenges related to knowledge caps of technology and component identification, mentioned in Section \ref{section:challengeslegacy}, could be solved.

\end{itemize}

\subsection{LLMs Used for Evaluation}
\label{evaluated:llms}

As shown in Table \ref{tab:usedmodels}, we selected five different LLMs for evaluation. The table includes the company, model, context length, and categorized size. We chose the latest LLMs to ensure that our evaluation was based on the most current and state-of-the-art models available for code generation. 

\usedLLMs

\subsection{Verification Process}
\label{verification}

As in \cite{enase25}, VAPU was verified by comparing it's results to OSL/ZSL prompts in different updating tasks in a legacy web application files. The ZSL prompts do not include any examples, while an OSL prompt includes one example \cite{brown_language_2020}. The files were updated with a ZSL prompt by default. If an example was required to guide the model’s behavior, an OSL prompt was used. The use of OSL/ZSL prompts as a comparing metric is supported by a previous study of the literature \cite{ouedraogo_large-scale_2024}, where OSL and
ZSL prompts were compared with different reasoning methods for LLMs. Each test was conducted with LLM's temperature first set at 1 and then to 0.

Each test was repeated in both methods ten times, taking into account the stochastic nature of LLM \cite{brown_language_2020}. This also included the zero temperature tests as the result was not deterministic in any of the evaluated LLMs. The loop in the system between the verifier agent and the finalizer agent was set with a maximum of two iterations. The verification process was conducted in two test types, which the first one focused on formed errors and the second part for the succeeded requirements of the outcome.

For every updated file tested with generated errors, static and dynamic testing was conducted to find all the resulting errors formed in the code. The errors were categorized into four types, which are defined below:

\begin{itemize}
\item[1.] \textit{Fatal Errors:} Errors that prevent a file from running, for example, syntax errors.
\item[2.] \textit{Runtime Errors:} Errors that do not prevent running the file, but during execution will prevent some functionality from it.
\item[3.] \textit{Content Errors:} The feature works, but its functionality is different and unintended from the original file.
\item[4.] \textit{Missing/Additional features:} The updated file lacks or includes additional features not existing and not intended in the file.
\end{itemize}

The errors were categorized into different types, as done in the literature, mainly focusing on separating fatal/compilation, runtime, and content/functional errors from each other \cite{dou_whats_2024,liu2024refining}. The same error caused by the same mistake was counted only once to ensure
that, for example, recurring syntax errors do not disproportionately affect the comparison with
other types of errors. Unlike in Ala-Salmi \textit{et al}. \cite{enase25}, a category for failed generations is not used. Reasoning for this was to get the equal number of analyzed files for all methods, as this category could happen due to outside factors such as a bad connection to an LLM API.

With files tested with the fulfilled requirements, the entire code was refactored and based on complex requirements. The evaluation process was ranked according
to the passed requirements. Every correctly working requirement resulted in value 1 and every incorrectly working requirement resulted in 0 value. The evaluation was conducted similarly to error counting with both static and dynamic test cases.

In addition to error or requirement counting, time duration and Lines of Code (LOC) were measured in most of the test cases. The
reason behind counting LOC was to analyze the difference of produced code length
between an OSL/ZSL prompt and VAPU. With time, a normalization was made to analyze how longer duration impacts VAPU's performance with errors and requirements in the tasks. From the results, we calculated Standard Deviation (SD) for the resulting errors and requirements to have a suitable metric to evaluate VAPU's performance.

\subsubsection{Used Material of the Verification Process}\hfill\\

The legacy application used in the verification process is called the Electronic Dictionary
Project (EDP) introduced by Norri \textit{et al}. \cite{norri_digitization_2020}. EDP contains a medieval English medical dictionary, with means to examine and modify an PostgreSQL database, which contains the data of the dictionary. This meant that EDP is a research tool for the mentioned dictionary. EDP is built on CakePHP, which is a PHP-based web framework. The version EDP uses is a legacy version 1.2 from 2008 \cite{koschuetzki_extra_2008}. In this study, the task was to update the version to 4.5 from 2023 \cite{cakephp_cakephp_2023}, making the version cap 15 years. 

CakePHP is built based on the Model-View-Controller (MVC) architecture \cite{cakephp_introduction_2022},
which is a software design pattern in which the view is the user side, model the database, and the controller acts as a mediator between
a model and a view. Between the two versions of CakePHP. Both the design and the syntax of the web framework have changed. One big change was CakePHP 3.0 in which Object–Relational
Mapping (ORM) was re-built \cite{cakephp_30_2024}. An another example is that the required version of PHP was raised to 7.4 in version 4 of CakePHP \cite{cakephp_installation_2024}, which was between 4 and 5 in the legacy version \cite{cakephp_introduction_2022}. As such, the update needs to take account of multiple different changes simultaneously to produce the correct code.

The files used from EDP are shown in Table \ref{tab:evaluated_files}. The list consists of six view files with varying LOC and different challenges. The views belong to two components in EDP with Views B-F forming a whole component in the application. The purpose of files are related, for example, for finding medieval English variants and related information such as name and original language \cite{norri_digitization_2020}. Views A-D were used in the error counting part of the verification process, and Views E-F with the requirements part of the process.

\usedfiles

View A is a simple reference list that shows a reference and terms related to a certain reference identifier. Due to it's simple structure it was chosen as the first view in the verification process. The challenge
related to the file update was a change in access to the data with a different name. View B lists quotes that are connected to a specific variant with an option of a certain origin language. View C, on the other hand, is a search form for finding variants in the database. Views B and C have a changed need to use arrays due to ORM ResultSets introduced in CakePHP 3 \cite{cakephp_retrieving_2024}.

View D is a dynamic dictionary that shows variants based on their first letter and with the option to filter using the language of origin. It contains both PHP and JavaScript code and was recognized as a challenge in the updating process. Due to two programming languages and higher LOC View D is the most challenging view in the error counting part of the verifying process.

Views E and f are both element view files that provide functions to View C search form. View E is an Ajax form used in View C that needed to be remade with the jQuery JavaScript library. As such, both the architecture and CakePHP code needed both to be updated concurrently. View F is a highlighting feature in View C that highlights the search world in the search results. Besides of the highlight functionalities, four helper functions making the highlight possible were included to be replaced with modern library alternatives. The definitions of the expected requirements were defined according to the verification process for both views and are listed below:

\begin{takeawaybox}[Requirements for View E form update and View F highlight update]
\scriptsize
\textbf{Requirements of the updated form of View E:}
\begin{itemize}
  \item[$\bullet$] The form sends data and receives results as expected.
  \item[$\bullet$] Dropdown menus show correct suggestions.
  \item[$\bullet$] The send button and dropdown objects work as intended.
\end{itemize}
\textbf{Requirements of the highlighting feature of View F:}
\begin{itemize}
  \item[$\bullet$] The highlighting works in normal case with no special characters or wildcards.
  \item[$\bullet$] Highlighting words with special medieval English letters.
  \item[$\bullet$] Highlighting works with wildcards with the highlight only on the part of the word where the wildcard was placed in the search query.
\end{itemize}
\end{takeawaybox}

For View E, a jQuery file was accepted as a URL or a local file. Also, a data name required by the controller was different and, therefore, disregarded from the requirement evaluation. The results of View F were measured along with the
number of correctly replaced functions using modern library implementations. Remaining functions were used in the file
with added guard to not re-declare them in the verification process. Besides these, no additional operations were added to the files in the evaluation. 

The prompts used in the evaluation process for both methods are shown in Table \ref{tab:usedprompts}. Compared to the used prompts in Ala-Salmi \textit{et al}. \cite{enase25}, a couple of changes were made. The ZSL prompt originally used in View B is disregarded due to the better performance of an OSL prompt which is used solely in the comparison. Similarly, View F only uses division to two tasks instead of one to make a clear comparison between VAPU and the ZSL method. The last sentence in paragraphs for View B and C prompt was only used in View B as it's requirement does not exist for View C. View D prompt could include text: "Only view side is returned as it was given." due to a faulty returning format.

\usedpromptsl

\subsection{Validation Process}
\label{validation}

The validation process was conducted with 20 Python projects with varying updating tasks from MIT-licensed GitHub repositories. As in the verification process, the amount of LLMs remained the same and a similar comparison to ZSL prompts was made. The difference was that the code was updated as Pass@1 for each file and LLM to validate the real use of the system. In addition, the temperature was selected based on the results of the verification process by taking into account both the resulting errors and the correct requirements.

Python projects were selected based on three features. The first feature was LOC which has been shown to have impact with an LLM's capacity to provide correct code \cite{dou_whats_2024,liu2024refining}. The second feature was the code complexity calculated using
Radon, a Python metrics tool \cite{Radon} which was used to calculate the Cyclomatic Complexity (CC) of the code as a letter from A to F. The code complexity is also found to increase LLM's difficulty to produce the correct code \cite{liu2024refining}. The last feature was the task amount for each update, which was expected to pose an additional challenge in having to conduct multiple tasks at once on the code.

Each selected project file was given an estimated difficulty value. For each rounded 100 LOC, the highest CC level found in the code, and each task, an additional point was given. The total difficulty was decided to remain between 3-10 with different projects equally distributed across it. The tasks were related to different updating requirements, such as changing used libraries, implementing user interface from console, or extending some feature based on new needs. The tasks and the used prompts are available in a dataset \cite{tampere_university_2025_16790515}.

The selected projects are shown in Table \ref{tab:github_used_files}. Table \ref{tab:github_used_files} shows the project with a reference to the repository. This is followed by LOC, CC, the amount of tasks, and the estimated difficulty of the update operation. Each updated file was evaluated based on three check marks which each updated file could earn based on different requirements. The first check mark was earned if the file included all the necessary updates and the code looked correct. The second check mark was earned if the implementation had no errors in the usage of the basic functions. The last check mark was earned if all the updated requirements were fully correct.

\githubfiles

\section{Results}
\label{Results}
In this section, we present the results of our proposed system. We first tested the reliability and effectiveness of the proposed system with six EDP view files whose results are provided in Section \ref{Suitability RQ1}. Then, we validated the results with 20 Python projects whose results are provided in Section \ref{Validation RQ2}.

\subsection{Reliability and Effectiveness of the Proposed System for Updating Deprecated Files (RQ)}\label{Suitability RQ1}

Views A-D were updated according to the verification process by counting every different error from the resulted updated view files. The results for OSL/ZSL prompts are presented in Table \ref{tab:results} and the results for VAPU in Table \ref{tab:results2} at both temperatures. Both tables are shown as the average of the ten runs done for each testing case. The tables include the average of errors for individual files which is followed by the average total and it's standard deviation. The LLMs are ordered by their estimated parameter size in ascending order.

Table \ref{tab:results} shows the results of OSL/ZSL prompts with an average of 2.02 and 2.04 errors depending on the temperature. When this is compared to the results of VAPU in Table \ref{tab:results2}, the rate on average is higher with 3.30 errors, which is reduced to 2.30 errors with the temperature set at 0. The standard derivation of errors is lower in both methods but more with OSL/ZSL prompts with no derivation observed with some LLMs. The time for VAPU to update four files is around four minutes longer due to more inference calls needed by the system.

\errorcountresultstz

In individual views, Views A and D had better results with a ZSL prompt, View B was tied, and View C had a lower error rate with VAPU. The error rate increased with a higher LOC in both methods, which was highest with View D in all cases. Overall, the results observed indicate that in a lower temperature VAPU can have a similar rate of error as OSL/ZSL files when updating legacy files to modern versions.

\errorcountresultstf

\begin{figure*}[h]
\centering
\includegraphics[scale=0.70,alt={Bar chart comparing error types between OSL/ZSL prompts and VAPU for each LLM. The chart shows similar error amounts between methods, with VAPU having fatal errors more common and runtime errors for OSL/ZSL prompts. Content and missing content errors are rare in every bar.}]{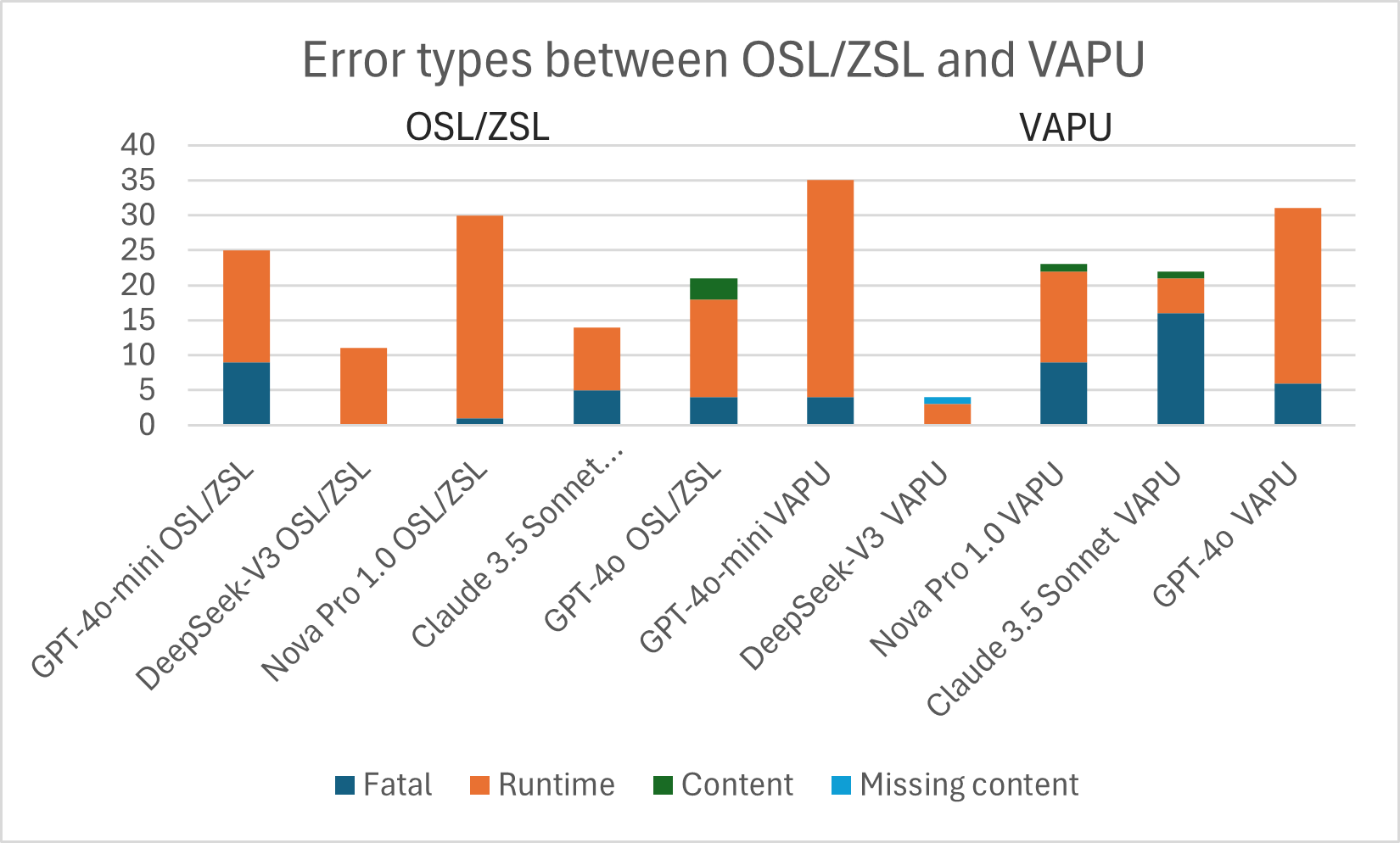}
\caption{Error types of OSL/ZSL prompts at T=1 and VAPU at T=0}
\label{fig:fig_errortypes}
\end{figure*}

The total error for Views A-D was 101 when using OSL/ZSL prompts with a temperature of 1, and 102 when the temperature was set to 0. In VAPU, the error amount was 165 with temperature set to 1 and 115 when temperature was at 0. In Figure \ref{fig:fig_errortypes} the lowest error rate is shown for both methods listing the four different error types for each LLM used.

As seen from Figure \ref{fig:fig_errortypes}, the majority of produced errors where runtime errors especially with OSL/ZSL prompts. Runtime errors included errors such as wrong URL links or faulty JavaScript elements. Fatal errors could also be the most common type of error with VAPU. The fatal errors where mostly related to hallucinations such as referring to non existent method in CakePHP or in rare cases syntax errors in the code. Content and missing content errors were relatively uncommon in updating tasks, but where possible especially longer code files such as View D.

The error rates between LLMs varied, with DeepSeek-V3 particularly doing well without a single fatal error in test cases. With VAPU the amount of all errors was only four, tied only with OSL/ZSL and Nova Pro 1.0 with the temperature set to 0. The medium-sized LLMs had overall the lowest rate of errors with the small parameter LLM and the large parameter models producing higher number of errors.

\begin{figure*}[h]
\centering
\includegraphics[scale=0.68,alt={Dot plot and linear regression of the error rate of VAPU in temperatures 0 and 1, with time normalized. The linear regression line when temperature was set to one is almost vertical, and when the temperature was set to 0, there is a slight increase with time. Most dots are between zero and three errors, with a range of -2.00 and 3.00 of normalized time.}]{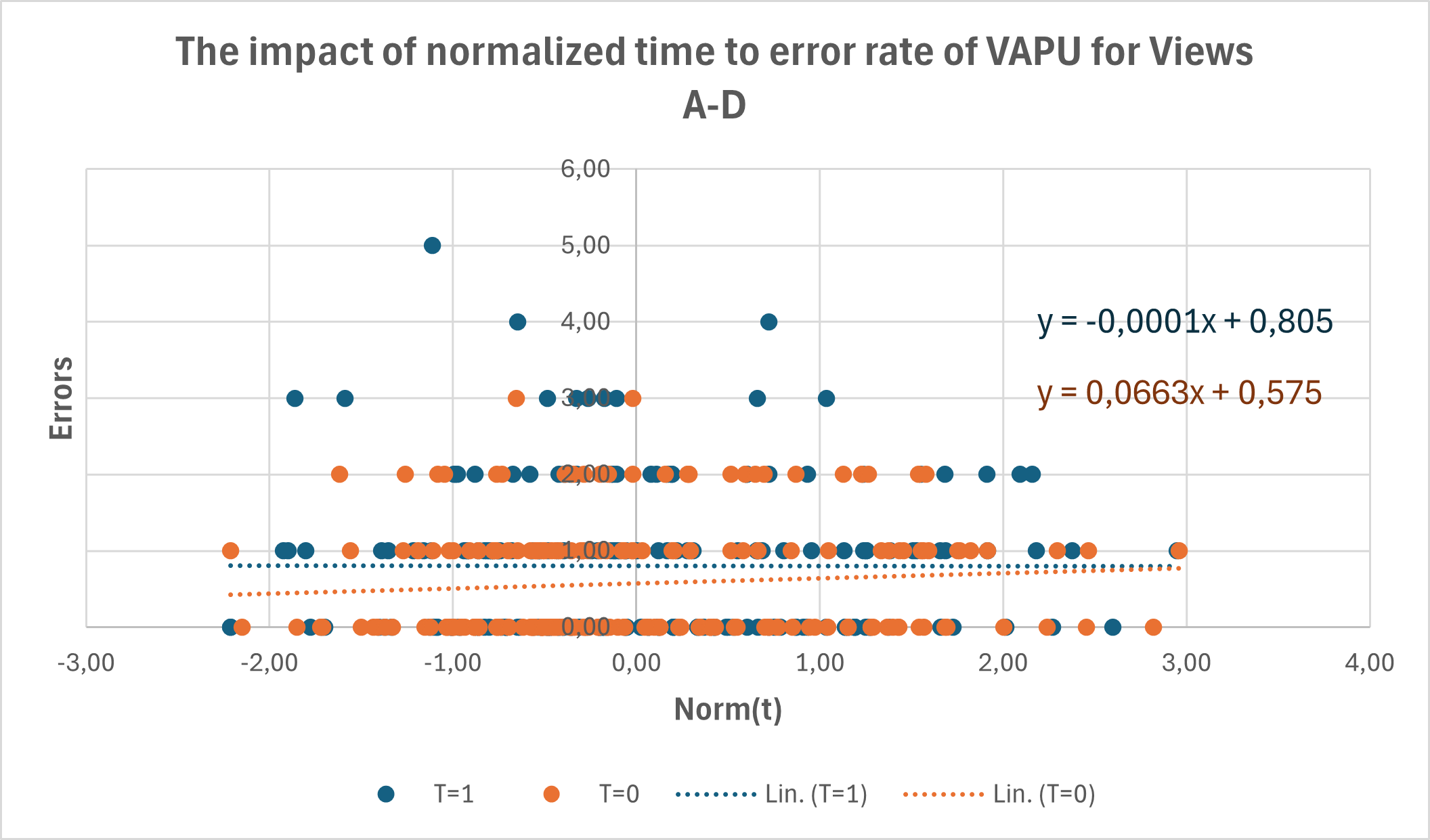}
\caption{Normalized time's impact on VAPU's error rate for Views A-D}
\label{fig:fig_normalized}
\end{figure*}

As explained in the methodology, the time was normalized with View A-D update results and compared to the resulted errors. The resulting dot plot is shown in Figure \ref{fig:fig_normalized} with both temperatures used for VAPU. Based on the plot, a linear regression was calculated for both temperatures showing no change with the temperature set to 1 and a slight increase when it was set to 0. With the temperature set at 0 the error rate was lower in low time durations but increased to similar levels with increased relative time. The results indicate that a longer duration with the verifier and the finalizer agent does not seem to improve the error rate further with more time.

View E was updated to a new form with three possible requirements to fulfill from the outcome. The results of each requirement are visible for the ZSL prompts in Table \ref{tab:results3} and for VAPU in Table \ref{tab:results4}. Similar to Views A-D, the results are shown as the average of the ten update attempts. The first value shows the result when temperature was set at 1 and the second value when it was set at 0. The average fulfilled requirements for ZSL prompts was 0.88 and with temperature lowered 1.36. For VAPU the same values were 0.64 and 0.96. As such, a ZSL prompt was slightly better on average in this view.

\requirementresultsE

In individual requirements, each requirement had a lower rate of success than a previous one. With the first requirement had at best 80\% success rate with ZSL and 50\% rate with VAPU. The last one, however, only had at best 10\% success rate with ZSL and 6\% success rate with VAPU. ALOC was higher with VAPU with around 40\% more code produced. This included, for example, navigation, style, or security features.

\requirementresultsEV

The update of View F was done similarly, with the exception of counting the correctly replaced helper functions from the updated files. The result of the ZSL prompt is shown in Table \ref{tab:results5} and VAPU in \ref{tab:results6}. In this view, VAPU was slightly better with 0.74 and 1.50 average completed requirements depending on the temperature. With a ZSL prompt this was 1.20 and 1.14 with the lower temperature.

\requirementresultsF

VAPU had the highest rate of 58\% and 56\% change of success in the first two requirements. With the last, ZSL had a 2\% higher success rate of 38\%. With ZSL only higher parameter models started to converge towards success, while with VAPU this included the lower parameter models in low temperatures. The replaced functions were more successful with OSL/ZSL prompts with around two more fully replaced functions on average. ALOC was higher in VAPU mainly due to lesser functions replaced in the updated files.

\requirementresultsFV

When Views E and F are taken into account, VAPU shows slightly better performance with View F on average with an expense of lesser replaced functions. One notable remark with VAPU was that Claude 3.5 Sonnet was incompatible with these tests due to unexpectedly larger code generation which consisted of some unnecessary code. This highlights the fact that the verification process might have highly varying results between LLMs depending on how compatible LLM is to an architecture of a multi-agent system. As such, verification with each LLM is required separately due to differences among LLMs. 

\noindent
\textbf{Key findings:} Overall, the verification process has shown, that VAPU is capable of updating legacy application view files with a similar range of error and fulfilled requirements as one inference call with an OSL/ZSL prompt. As such, a multi-agent system is able to modernize deprecated code with the limitations of errors and fully correct requirements. Lowering of the temperature improved the results both in the error rate and the fulfilled requirements using VAPU. The error rate had an improvement of 33\% and the fulfilled requirement had 80\% when the average values were compared to each other.

\subsection{Validation of the Proposed System for Updating Deprecated Files (RQ)}
\label{Validation RQ2}

As defined in the validation process, the temperature used for the validation for each method is the combination of the error rate and the fulfilled requirements from the verification process as $T^u=AR-\frac{1}{2}AE$. Based on the results, all methods except GPT-4o-mini with ZSL prompts were conducted with the temperature set to 0 due to better performance.

The results of the VAPU and the comparison with the ZSL prompts are visible in Table \ref{tab:gitlab_results2}. The table lists each check mark obtained by VAPU in each LLM in every project. If the result was not same with ZSL the result includes an overscore to indicate different result. The total check marks for each LLM are visible for both VAPU and ZSL at the bottom of the table. Values marked with $1\atop$ have been conducted with shortened code file due to the limitations of the LLM output size in certain models.

\pythonresulttableVAPU

The best performing model for ZSL was Claude 3.5 Sonnet, similar to the requirement section of the verification process. Claude 3.5 Sonnet obtained 46 check marks, converting to a pass rate of 76.6\%, which was the highest observed rate in this study. Three models GPT-4o-mini, GPT-4o, and DeepSeek-V3 all reach a range of 37-38 check marks. Nova Pro 1.0 scored the lowest value of 31 check marks of the compared LLMs.

For VAPU, the obtained check marks are in similar range than ZSL prompts in three models and notable better with GPT-4o and Nova Pro 1.0 with each ranked seven check marks higher than the ZSL prompts. This change lifted GPT-4o to the same range as Claude 3.5 Sonnet and Nova Pro 1.0 to the same range as GPT-4o-mini and DeepSeek-V3. The values correspond to the LLM size, with the two large parameter model having now the leading result while the three lower parameter models have around the same performance.

When it comes to the different check mark types, ZSL had 81 first check marks, 75 second check marks, and 33 third check marks, with a total of 189 check marks. VAPU on the other hand had 94 first check marks, 70 second check marks, and 40 third check marks, with a total of 204 check marks.

The check marks obtained by VAPU by code length, code complexity, and task amount are visible in Table \ref{tab:checkmarksbytype}. The table compares the obtained check marks to ZSL prompts by showing the difference in brackets. In each row, the best performed LLM is in bold. The results show that code files that have either higher LOC, cyclomatic complexity or higher task amount, VAPU might increase the possibility of correct result in some LLMs. For LOC the check mark difference is around equal under 300 LOC but rises to +12 for VAPU when all LLMs are taken into account.
With cyclomatic complexity the task rate changes from +11 to -5 and then to +10, which could be explained by the complexity becoming much higher in the last category. With tasks, the task amount is again equal with only one task, but increases to +12 with two tasks and +5 with three tasks. 

\pythoncomparisonVAPU

\noindent
\textbf{Key findings:} Based on the results of the validation process, VAPU is capable of improve the probability of correct code in certain LLMs when code is updated. The improvement focuses mainly on longer code files and complex code, as seen in Table \ref{tab:checkmarksbytype}. The rate of improvement in this metric in pass@1 was 18.4\% and 22.6\% in the LLMs that were compatible in this implementation.

\section{Discussion}
\label{Discussion}

In this study, we proposed a multi-agent system VAPU to update legacy applications in a cost-effective way to answer up to 95\% of the deprecation rate of web applications \cite{demir2021our}. The results indicate that the multi-agent system is capable of updating legacy files with a similar error rate as OSL/ZSL prompts with a higher succeed rates observed in some updating tasks. The results of VAPU improved at the lower temperature and with LLMs compatible with its prompting. High-parameter LLMs GPT-4o and Claude 3.5 Sonnet did excel in harder tasks with specified requirements while lower parameter models DeepSeek-V3 and Nova Pro 1.0 could obtain the lowest error count in version update tasks.

When the success rate of the updated Python files is compared to the HumanEval metric and the LLM-based systems studied in Table \ref{tab:humaneval}, the baseline improvement of VAPU with compatible LLMs was in a similar range as the improvement in HumanEval using the alternative academic systems. The improvement of 18.4\% and 22.6\% indicates further that a multi-agent system could improve the probability of correct code in real projects when the code is updated instead of created only from a description. When CC and LOC were measured to the expected performance, both the VAPU and OSL/ZSL prompts had a degrease in quality as also found in the literature \cite{dou_whats_2024,liu2024refining}. With the task amount, VAPU had a slight improvement compared to ZSL prompts. The good results with ZSL prompts handling multiple tasks are also supported by the literature \cite{son2024multi}. This might suggest that a multi-agent system could also do tasks in the same inference call.

Multi-agent systems being able to update existing code have opportunities in the software industry. In practitioners perspective, software engineers desire to use LLMs for code improvement and maintenance to automate these processes \cite{russo2024navigating}. One of the reasons was the reparative nature of tasks, which, as stated earlier, take 40-90\% of a company's resources \cite{davis_97_2009,erlikh_leveraging_2000,russo2024navigating}. The possibility of automatizing the maintenance process by at least partially with LLMs will result in a faster and cheaper process for companies. As such, by providing benefits with both software engineers and companies, the adoption of these systems for code maintenance is a possibility. What VAPU can provide is a foundation for further refined systems for practitioners that can provide the desired legacy application update in a cost-effective manner.

For academic perspective, the study provides multiple opportunities. The improved performance at lower temperature might be repeatable with other systems and models using tokens inside the system/model to build the outcome. This could include reasoning models which are using up to thousands of reasoning tokens to produce the answer \cite{guo2025deepseek}. Another finding was that medium-sized LLMs excelled in the error amount and large-sized LLMs excelled in more challenging requirement tasks in the verification process. Similarly in the validation process, VAPU had lowest relative performance with low-difficulty tasks compared to ZSL prompts. These findings correspond to finding that easier tasks are better solved by LLM's while the more complex reasoning models perform overall better in more difficult problem solving \cite{shojaee2025illusion}. As such, this finding could support the idea that more complex generative AI models and systems could underperform in less complex tasks.

In future work, more advanced multi-agent systems could be explored with larger and more challenging coding update tasks. The advanced version of VAPU could include a better verifier with testing tools or reasoning abilities. A larger scale update consisting of a whole project could be a natural step with investigating real benefits of an LLM based multi-agent system. Additionally, the impact of the temperature parameter for the multi-agent performance overall could warrant additional investigation.

\subsection{Challenges and Limitations of the Study}

Overall, despite relatively good results, challenges were observed related to VAPU:

\textbf{Agentic prompt chain degradation:} The better results of VAPU at lower temperatures suggest that the higher variation in the answers will likely make VAPU fail to produce the correct code. As an increase of the temperature makes more unlikely answers more probable \cite{KoulouriotisD.E.2008Rlae}, the multi-agent system seems to be more prone to the errors resulting from it than an OSL/ZSL prompt. This is similar to a telephone game, in which information is passed in a chain from one player
to another. The longer the game continues more distorted the original message becomes. With a system of multiple agents, this seems to be a challenge that needs to be addressed with more guardrails inside the system. 

\textbf{Verifier functionality:} As found in the time analysis, the error rate did not improve with a longer time duration of the system. This indicates that the verifier needs either a better model or tools to provide more correct answers. A debugger with test cases could help to provide more correct results from the verifier agent \cite{huang_agentcoder_2024,zhong_debug_2024}. An alternative option is a reasoning model that is capable of more challenging problem solving and could correctly identify errors more likely while not fully correct \cite{shojaee2025illusion}. 

\textbf{LLM capability:} The capability of LLM is a limiting factor as seen in the lower results of smaller-parameter LLMs. As the parameter size limits the capabilities of LLM \cite{kaplan_scaling_2020}, this will also limit the probability that an agent performs correctly. To improve agents overall, for example, better prompt engineering techniques could improve the agent's even with lower parameter models \cite{white2023prompt}. In an alternative way, a reasoning model could take the agent's role as suggested for the verifier agent.

In addition, this study recognizes limitations to the results. The verification process of the system was conducted only with six view files and an additional number of view files with different features could give different results due to the stochastic nature of LLMs \cite{brown_language_2020}. This could also apply if more LLMs or more temperature values were included in the process. With the validation process, more files and tasks could
give similarly different outcomes. The exact impacts of code length, CC, and the number
of tasks could benefit for individual
testing to give a more specific information of each parameter's impact to the outcome.

There are also limitations with the validity of the proposed system. A possible error in the system could
give weaker results, for example, if one of the agents
has poor instructions to operate. This is also possible with the prompts, which could be better with different prompt engineering techniques, giving better results with both VAPU and OSL/ZSL prompts. With the time analysis, the varying service time of API based LLMs could have an impact on the results. 

\section{Conclusion}
\label{Conclusion}

To solve the problem of widespread amount of legacy applications online, we proposed a multi-agent system VAPU to update them cost-effectively. In this study, we extended VAPU's evaluation from one LLM to five, tested the functionality of system in lower temperature, and validated the system with 20 Python GitHub project files. The results indicates that updating legacy code is possible with multi-agent systems, which can afford companies to conduct code maintenance partially automated. This would reduce the number of deprecated applications, improving the security, reliability, and compatibility of applications, reducing the risks for companies. For academia, the better results of VAPU in lower temperatures and the benefits focusing mainly on complex tasks give insight of the contrast between a system using multiple inference calls to produce results and one inference call prompts such as ZSL/OSL prompts.

Using a multi-agent system to automate code maintenance and reducing the amount of legacy applications shows a promise and more research is needed for the area. Challenges related to long agent chains, verifier, and overall LLMs were observed which require more solutions. VAPU provides a functioning foundation to provide solutions to these challenges. For a future system, a large-scale application update is a next step forward in code maintenance automation.

\bibliographystyle{splncs04}
\bibliography{references}

\end{document}

%% file: tables/results.tex
\newcommand{\pythonresulttableVAPU}{

\begin{table}[h]
  \small
  \begin{center}
    \caption{Results of GitLab updates with VAPU compared to ZSL}
    \label{tab:gitlab_results2}
    \begin{tabular}{l | P{1.5cm} P{1.5cm} P{1.5cm} P{1.5cm} P{1.1cm} }
     \hline

      \textbf{Project}  & \textbf{GPT-4o-mini}  &\textbf{GPT-4o} & \textbf{DeepSeek -V3}& \textbf{Claude 3.5 Sonnet}& \textbf{Nova Pro 1.0}  \\
                     
      \hline

      Address-Book &$\checkmark \overline{X}\overline{X} $&$\checkmark\checkmark \overline{X} $&$\checkmark\checkmark\overline{\checkmark} $&$\checkmark\checkmark \overline{X}$&$\checkmark\checkmark\overline{X}$\\

      Alarm clock &$\checkmark\overline{\checkmark}\checkmark$&$\checkmark \overline{X}\checkmark $&$\checkmark\checkmark\checkmark $&$ \checkmark\checkmark\checkmark$&$\checkmark\overline{\checkmark}\checkmark $\\

      Billing system &$\checkmark X X$&$\overline{\checkmark}\overline{\checkmark} X$&$\overline{\checkmark X} X ^{1}$&$\overline{ X X} X ^{1}$&$\overline{\checkmark}\checkmark X^{1}$\\

      Bouncing ball simulator &$\checkmark\checkmark X$&$\checkmark\checkmark\checkmark$  &$\checkmark\checkmark\checkmark$ &$\checkmark\checkmark\overline{X}$&$\checkmark \overline{X} X $\\

      Cafe management syst.&$\overline{\checkmark\checkmark} X$&$\overline{\checkmark} X\overline{\checkmark}$&$ X X X$&$\checkmark\checkmark \overline{X}$&$\overline{\checkmark} X X$\\

       Finance Tracker&$\checkmark X X$&$\checkmark\checkmark\overline{\checkmark}$&$\checkmark X X$&$\checkmark\overline{\checkmark\checkmark}$&$\checkmark X X$\\

       Geometry &$\checkmark\checkmark X$&$\checkmark\checkmark X$&$\checkmark\checkmark X$&$\checkmark \overline{X} X$&$\overline{\checkmark\checkmark\checkmark}$\\

        Health Log Book &$ X\checkmark X$&$\checkmark\checkmark X$&$\checkmark \overline{X} X$&$\checkmark \overline{X\checkmark}$&$\checkmark \overline{X} X$\\

         Lazy Pong &$\overline{\checkmark X} X$&$\overline{\checkmark X} X$&$ X \overline{X} X
         $&$\overline{\checkmark}\checkmark X$&$ X\checkmark X ^{1}$\\

          Notepad &$\checkmark X X$&$\overline{\checkmark\checkmark} X$&$\overline{\checkmark} X X$&$\checkmark\checkmark X$&$\overline{\checkmark} X X$\\

          Organize Directory &$  \checkmark\checkmark \overline{X}$&$ \checkmark\checkmark\checkmark $&$ \checkmark\checkmark\overline{\checkmark}$&$\checkmark\checkmark\checkmark $&$\checkmark X\checkmark$ \\

          Password &$ \checkmark\checkmark\checkmark$&$ \checkmark\checkmark X $&$\checkmark\checkmark \overline{X} $&$ \checkmark\checkmark X$&$\checkmark\overline{\checkmark X}$ \\

          Python-snake &$\checkmark X X $&$\checkmark X\overline{\checkmark}  $&$ \checkmark \overline{X} X  $&$ \checkmark\checkmark\checkmark $&$\checkmark\checkmark X$ \\

          Receipt generator &$\checkmark\checkmark X $&$ \checkmark\checkmark X $&$  X\checkmark X$&$\checkmark\checkmark\checkmark$&$\overline{\checkmark X} X$ \\

           Restaurant-Management syst.&$\checkmark\checkmark\checkmark$&$\checkmark\checkmark X$&$\checkmark \overline{X\checkmark}$&$\checkmark\checkmark X$&$\checkmark\checkmark X$\\  

           Scientific-Calculator &$\checkmark\checkmark \overline{X}$&$\checkmark\checkmark X$&$\checkmark\overline{\checkmark\checkmark}$&$\checkmark\checkmark\overline{\checkmark}$ &$\checkmark\overline{\checkmark} X$\\

          Simple calculator &$ \checkmark\checkmark \overline{X} $&$ \checkmark\checkmark\checkmark $&$\checkmark \overline{X}\checkmark  $&$\checkmark\checkmark\checkmark $&$\checkmark\checkmark\checkmark$ \\

        Simple Http Server &$  \checkmark\checkmark\overline{\checkmark}$ &$\checkmark\checkmark\checkmark  $&$ \checkmark\checkmark\checkmark $&$\checkmark\checkmark\checkmark$&$\checkmark\checkmark\checkmark$ \\

         Search books by ISBN &$\checkmark\checkmark\overline{\checkmark} $&$\checkmark\checkmark X  $&$ \checkmark\checkmark\overline{\checkmark}$&$\checkmark\checkmark\overline{\checkmark}$&$\checkmark\overline{\checkmark\checkmark}$ \\

        Sudoku-Solve &$\checkmark\checkmark X$&$\checkmark\checkmark\overline{\checkmark}$&$\checkmark\checkmark\checkmark$&$\checkmark X X$&$\checkmark\checkmark X$\\

        \hline
        Total ZSL/VAPU &37/\textbf{38}&38/\textbf{45}&37/\textbf{38}&\textbf{46}/45&31/\textbf{38}\\
     
      \hline
    \end{tabular}
  \end{center}
\end{table}
}

\newcommand{\pythoncomparisonVAPU}{

\begin{table}[h]
  \small
  \begin{center}
    \caption{Check marks by CC, LOC and the task amount for VAPU}
    \label{tab:checkmarksbytype}
    \begin{tabular}{l | c| P{1.8cm} P{1.8cm} P{1.8cm} P{1.8cm} P{1.8cm} }
      \hline
        \textbf{Feature}& \textbf{Max} & \textbf{GPT-4o-mini}  &\textbf{GPT-4o} & \textbf{DeepSeek -V3}& \textbf{Claude 3.5 Sonnet}& \textbf{Nova Pro 1.0}  \\
      
      \textbf{LOC} & &  & & & &   \\

      \textbf{<100} &21&18 (+2)  &17 (-1) & 18 (+1)& \textbf{19 (-)}& 16 (+2)  \\

      \textbf{100-300} &18& 9 (-1)  &\textbf{13 (+1)} & 10 (-1)& \textbf{13 (-)}& 11 (+1)  \\

      \textbf{300>} &21& 11 (-) &\textbf{15 (+7)} & 10 (+1)& 13 (-1)& 12 (+4)  \\

      \textbf{CC} & &  & & & &  \\

      \textbf{A} &24  & 17 (+3)  &18 (+2) & 15 (+1)& \textbf{21 (+3)}& 15 (+2)  \\

      \textbf{B} &18 & 11 (-2)  &\textbf{14 (-)} & 13 (-2)& 13 (-3)& 13 (+2)  \\

      \textbf{C-E} &18 & 10 (-)&\textbf{13 (+5)} & 10 (+2) & 11 (-1)& 10 (+3)  \\

      \textbf{Tasks} & &   & & & &   \\\textbf{}

      \textbf{1} &18 & 13 (-)  &\textbf{15 (+2)} & 12 (-3)& \textbf{15 (-)}& 13 (-)  \\

      \textbf{2} &24 & 14 (-1)  &16 (+4) & 14 (+3)& \textbf{17 (+1)}& 14 (+4)  \\

      \textbf{3} &18  & 11 (+2)  &\textbf{14 (+1)} & 12 (+1)& 13 (-2)& 11 (+3)  \\

      \hline
    \end{tabular}
  \end{center}
\end{table}
}

\newcommand{\errorcountresultstf}{
\begin{table}[h]
  \small
  \begin{center}
    \caption{Average different errors by model and method with VAPU in temperatures 1 and 0}
    \label{tab:results2}
    \begin{tabular}{l | P{0.5cm} P{1.3cm} P{1.3cm} P{1.3cm} P{1.3cm} P{1.0cm} P{1.0cm} P{1.0cm}}
      \hline
      \textbf{Model} & \textbf{T} &\textbf{View A} & \textbf{View B}& \textbf{View C}  & \textbf{View D} & \textbf{Total} & \textbf{SD}& \textbf{t (s)}  \\
                     
      \hline
      
      GPT-4o-mini & 1 & 0.50 & 0.60 & 1.00 &2.30 & 4.40 & 1.200 & 252.5 \\

      DeepSeek-V3	 & 1 & 0.00 & 0.10 & 0.50& 1.10 &1.70 & 1.100 &300.3 \\

      Nova Pro 1.0 & 1 & 0.80 & 1.10 & 1.40 & 1.10 &4.40 & 1.855&365.9 \\

      Claude 3.5 Sonnet & 1 & 1.20 & 0.20 & 0.40& 1.50 & 3.30 & 0.943&308.4 \\

      GPT-4o & 1 & 0.50 & 0.20 & 0.30& 1.60 & 2.70 &1.345&175.8   \\

      \hline
        \textbf{Average} &1  & 0.60 & 0.44 & 0.72 &1.52 &3.30 &1.289&280.6\\
      \hline

      GPT-4o-mini & 0 & 0.20 & 0.30 & 1.30 &1.70 & 3.50 & 1.565 & 396.9\\

      DeepSeek-V3	 & 0 & 0.00 & 0.10 & 0.10& 0.20 & 0.40 & 0.490 &514.7 \\

      Nova Pro 1.0 & 0 & 0.10 & 0.50 & 0.60 & 1.10 &2.30 & 1.005&262.7 \\

      Claude 3.5 Sonnet & 0 & 1.00 & 0.20 & 0.00& 1.00 &2.20 &0.872&289.7 \\

      GPT-4o & 0 & 0.60 & 0.00 & 0.60& 1.90 &3.10 & 1.446&149.9  \\

      \hline
       \textbf{Average} &0  & 0.38 & 0.22 & 0.52 & 1.18 & 2.30 &1.076&322.8\\

       \hline

    \end{tabular}
  \end{center}
\end{table}

}

\newcommand{\errorcountresultstz}{
\begin{table}[h]
  \small
  \begin{center}
    \caption{Average different errors by model and method with OSL/ZSL prompts in temperatures 1 and 0}
    \label{tab:results}
    \begin{tabular}{l | P{0.5cm} P{1.3cm} P{1.3cm} P{1.3cm} P{1.3cm} P{1.0cm} P{1.0cm} P{1.0cm}}
      \hline
      \textbf{Model} & \textbf{T} &\textbf{View A} & \textbf{View B}& \textbf{View C}  & \textbf{View D} & \textbf{Total} & \textbf{SD} & \textbf{t (s)} \\
                     
      \hline

      GPT-4o-mini & 1 & 0.00 & 0.40 & 0.70 & 1.40 &  2.50 &  1.285 &30.8\\

       DeepSeek-V3		
      & 1 & 0.00 & 0.10 & 1.00 &0.00 & 1.10 & 0.300 &71.2 \\

      Nova Pro 1.0&1  & 0.10 & 1.00 &1.10 &0.80 & 3.00 & 0.894&35.2 \\

      Claude 3.5 Sonnet & 1 &0.40& 0.00 & 0.00 & 1.00 & 1.40 & 0.663& 49.3 \\

      GPT-4o & 1 & 0.40 & 0.00 & 0.30& 1.40 & 2.10 & 0.943 &35.9  \\

      \hline
       \textbf{Average} &1  & 0.18 & 0.30 & 0.62 & 0.92 & 2.02 &0.817&44.5\\
      \hline

      GPT-4o-mini & 0 & 0.00& 0.90 & 3.00 & 1.50 &  5.40 & 0.663 &48.7\\

      DeepSeek-V3		
      & 0 & 0.00 & 0.00 & 1.00 &0.00 &1.00 & 0.000 &163.9 \\

      Nova Pro 1.0&0  & 0.00 & 0.20&0.20 &0.00 &0.40 &0.490& 28.2\\

       Claude 3.5 Sonnet & 0 &0.40& 0.00 & 0.00 & 1.00 & 1.40& 0.490&43.3\\

      GPT-4o & 0 & 0.00 & 0.00 & 0.00& 2.00 &2.00 &0.000&41.0   \\

      \hline
       \textbf{Average} &0  & 0.08 & 0.22 & 0.84 & 0.90 & 2.04 &0.329&65.0\\      
      
       \hline

    \end{tabular}
  \end{center}
\end{table}

}

\newcommand{\requirementresultsE}{
\begin{table}[h]
  \small
  \begin{center}
    \caption{Average accepted requirements of View E with ZSL prompts}
    \label{tab:results3}
    \begin{tabular}{l | P{1.6cm} P{1.6cm} P{1.6cm} P{1.6cm} P{1.6cm}}
      \hline
      \textbf{Model}  &\textbf{Regt 1} &\textbf{Regt 2}&\textbf{Regt 3}&\textbf{Total}&\textbf{ALOC}\\
                     
      \hline

      GPT-4o-mini  &0.00/0.00&0.50/1.00&0.00/0.00& 0.50/1.00&44/41  \\

      DeepSeek-V3 &0.90/1.00&0.00/0.00&0.00/0.00& 0.90/1.00&54/60  \\

      Nova Pro 1.0  &0.70/1.00&0.10/0.00&0.00/0.00& 0.80/1.00&48/42 \\

      Claude 3.5 Sonnet &0.50/1.00&0.40/1.00&0.20/0.20 &1.10/2.20&70/65 \\

      GPT-4o &0.60/1.00&0.30/0.30&0.20/0.30& 1.10/1.60& 42/37 \\
      
      \hline
       \textbf{Average} &0.54/0.80&0.26/0.46&0.08/0.10& 0.88/1.36 &52/49 \\

      \hline
    \end{tabular}
  \end{center}
\end{table}

}

\newcommand{\requirementresultsEV}{
\begin{table}[h]
  \small
  \begin{center}
    \caption{Average accepted requirements of View E with VAPU}
    \label{tab:results4}
    \begin{tabular}{l |  P{1.6cm} P{1.6cm} P{1.6cm} P{1.6cm} P{1.6cm}}
      \hline
      \textbf{Model}  &\textbf{Regt 1} &\textbf{Regt 2}&\textbf{Regt 3}&\textbf{Total}&\textbf{ALOC}\\
                     
      \hline

      GPT-4o-mini &0.30/0.00&0.30/0.80&0.10/0.00& 0.70/0.80&66/53 \\

       DeepSeek-V3  &0.60/0.80&0.40/0.20&0.00/0.10& 1.00/1.10&61/63 \\

      Nova Pro 1.0  &0.50/0.70&0.00/0.10&0.00/0.10& 0.50/0.90&45/48 \\

      Claude 3.5 Sonnet  &0.00/0.10&0.00/0.00&0.00/0.00& 0.00/0.10&150/141 \\
      
    GPT-4o  &0.60/0.90&0.40/0.90&0.00/0.10&  1.00/1.90&42/46 \\

      \hline
       \textbf{Average} &0.40/0.50&0.22/0.40&0.02/0.06& 0.64/0.96&73/70\\
      \hline
 
    \end{tabular}
  \end{center}
\end{table}

}

\newcommand{\requirementresultsF}{

\begin{table}[h]
  \small
  \begin{center}
    \caption{Average accepted requirements of View F with ZSL prompts}
    \label{tab:results5}
    \begin{tabular}{l |  P{1.6cm} P{1.6cm} P{1.6cm} P{1.4cm} P{1.3cm} P{1.3cm}}
      \hline
      \textbf{Model} &\textbf{Regt 1} &\textbf{Regt 2}&\textbf{Regt 3}&\textbf{Total}&\textbf{ALOC}&\textbf{RF}\\
    \hline

      GPT-4o-mini  &0.70/0.00&0.60/0.00&0.50/0.00& 1.80/0.00&42/40&3.20/3.30  \\

      DeepSeek-V3 &0.00/0.00&0.00/0.00&0.00/0.00& 0.00/0.00&45/40&3.80/3.30  \\

      Nova Pro 1.0   &0.20/0.00&0.20/0.00&0.10/0.00& 0.50/0.00&49/52&2.90/2.90 \\

      Claude 3.5 Sonnet &0.80/1.00&0.80/1.00&0.00/1.00 &1.60/3.00&34/32&3.80/4.00 \\

      GPT-4o&0.80/0.90&0.80/0.90&0.50/0.90& 2.10/2.70& 44/72&3.10/1.30 \\
      
      \hline
       \textbf{Average} &0.50/0.38&0.48/0.38&0.22/0.38& 1.20/1.14 &52/46&3.36/2.50 \\

       \hline

      \hline
    \end{tabular}
  \end{center}
\end{table}

}

\newcommand{\requirementresultsFV}{

\begin{table}[h]
  \small
  \begin{center}
    \caption{Average accepted requirements of View F with VAPU}
    \label{tab:results6}
    \begin{tabular}{l |  P{1.6cm} P{1.6cm} P{1.6cm} P{1.4cm} P{1.3cm} P{1.3cm}}
      \hline
      \textbf{Model}  &\textbf{Regt 1} &\textbf{Regt 2}&\textbf{Regt 3}&\textbf{Total}&\textbf{ALOC}&\textbf{RF}\\
                     
      \hline

      GPT-4o-mini &0.30/0.50&0.00/0.50&0.20/0.40& 0.50/1.40&52/68&2.70/0.50  \\

      DeepSeek-V3 &0.50/0.50&0.50/0.50&0.10/0.40& 1.10/1.30&82/58&1.30/2.90  \\

      Nova Pro 1.0  &0.40/0.50&0.40/0.40&0.10/0.10& 0.90/1.00&59/47&2.70/3.20 \\

      Claude 3.5 Sonnet &0.20/0.40&0.20/0.40&0.20/0.10 &0.60/0.90&113/87&0.60/1.60 \\

      GPT-4o &0.20/1.00&0.20/1.00&0.20/0.90& 0.60/2.90& 73/74&1.40/1.60 \\
      
      \hline
       \textbf{Average} &0.32/0.58&0.26/0.56&0.16/0.36& 0.74/1.50 &76/67&1.74/1.96 \\

      \hline
    \end{tabular}
  \end{center}
\end{table}

}

%% file: tables/methodology.tex
\newcommand{\usedpromptsl}{
\begin{table}[tb]
  \centering
  \small
  \caption{Used prompts in each view, adapted from \cite{enase25}}
  \label{tab:usedprompts}
  \begin{tabular}{>{\raggedright\arraybackslash}p{1.5cm}>{\centering\arraybackslash}p{2cm}>{\raggedright\arraybackslash}p{8.5cm}}
    \hline
    \textbf{File} & \textbf{Method} & \textbf{Prompt} \\
    \hline
    View A & VAPU/ZSL & "Update whole cakePHP view file from version
1.2 to version 4.5. Change \$mainreference
to \$bookReference while removing [’Reference’] between wanted objects, access it ORM
style. Write \$term lowercase." \\

    View B, C & OSL & "Update CakePHP view file from version 1.2
to version 4.5. Requirement: \$quotes are
ORM \textbackslash ResultSets made of arrays accessed
with [’Fieldname’][’fieldname’] and must be
accessed so in the updated file. Example: old
syntax: [’apple’][’lemon’] new syntax: [’Apple’][’lemon’]. (Use function first() instead of
[0] to the ORM object.)"  \\
    View B, C & VAPU & "Update CakePHP view file from version 1.2
to version 4.5. Requirement: \$quotes are
ORM \textbackslash ResultSets made of arrays accessed
with [’Fieldname’][’fieldname’] and must be
accessed so in the updated file. Example: old
syntax: [’apple’][’lemon’] new syntax: [’Apple’][’lemon’]. (Use function first() instead of
[0] to the ORM object.)" \\

View D & VAPU/ZSL & "Update whole cakePHP view file from version 1.2 to version 4.5. Use ORM access with
\$variant with direct access to name and id." \\ 

View E & VAPU/ZSL & "Update CakePHP version 1.2 ajax form into
CakePHP 4.5 version with jQuery architecture including jquery-3.6.0.min file. Make the
jQuery implementation fully functional version with dropdown updated every time the
letter is written to it" \\ 

View F & VAPU/ZSL & "Update CakePHP element view file from 1.2
to 4.5. Replace functions in the element view
file by ready made PHP libraries or update
them if not found" \\ 

    \hline
  \end{tabular}
\end{table}
}

\newcommand{\githubfiles}{

\begin{table}[h]
  \small
  \begin{center}
    \caption{Used GitLab projects in the validation process}
    \label{tab:github_used_files}
    \begin{tabular}{l | P{1cm} P{1cm} P{1cm} P{2cm}}
      \hline
      \textbf{Project}  & \textbf{LOC}  &\textbf{CC} & \textbf{Tasks}& \textbf{Difficulty} \\
                     
      \hline

      Address-Book \cite{GitHub7} &259&B&3&8\\

      Alarm clock \cite{GitHub4} &86&A&3&5\\

       Billing system \cite{GitHub4}&428&D&2&10\\

      Bouncing ball simulator \cite{GitHub4} &161&A&3&6\\

      Cafe management syst. \cite{GitHub9} &397&C&3&10\\

      Finance Tracker \cite{GitHub5} &161&A&3&6\\

      Geometry \cite{GitHub2} &288&B&3&8\\

      Health Log Book \cite{GitHub7} &259&A&2&6\\

      Lazy Pong \cite{GitHub5} &436&C&2&9\\

      Notebad \cite{GitHub9} &360&A&2&7\\

      Organize Directory \cite{GitHub6} &90&D&2&7\\

      Password  \cite{GitHub2} &97&B&1&4\\

      Python-snake  \cite{GitHub3} &195&B&1&5\\

       Receipt generator \cite{GitHub6} &97&A&1&3\\

       Restaurant-Management syst. \cite{GitHub7} &226&A&2&5\\

       Scientific-Calculator \cite{GitHub6} &378&C&2&9\\

       Simple calculator \cite{GitHub1} &342&B&1&7\\

       Simple Http server \cite{GitHub5} &64&A&1&3\\

       Search books by ISBN \cite{GitHub2} &82&A&2&4\\

       Sudoku-Solve \cite{GitHub8} &316&E&1&9\\

      \hline
    \end{tabular}
  \end{center}
\end{table}
}

\newcommand{\usedLLMs}{
\begin{table}[h]
  \small
  \begin{center}
    \caption{Selected LLMs for the evaluation}
    \label{tab:usedmodels}
    \begin{tabular}{l | c P{2.5cm} P{1.7cm}  c}
      \hline
      \textbf{Company} &\textbf{Model} & \textbf{Parameter size} \cite{abacha2024medec,liu2024deepseek} & \textbf{Context length} \cite{Openrouter} & \textbf{Size} \\
                     
    \hline

      Anthropic&Claude 3.5 Sonnet & \char`\~ 175B & 200K & L \\

      Amazon&Nova Pro 1.0& \char`\~ 40B & 300K		& M \\

      DeepSeek&DeepSeek-V3		
      & 37B (671B)& 128K		& M \\

      OpenAI&GPT-4o-mini & \char`\~ 8B & 128K & S \\

      OpenAI&GPT-4o & \char`\~ 200B& 128K & L \\

      \hline
    \end{tabular}
  \end{center}
\end{table}
}

\newcommand{\usedfiles}{
\begin{table}[htbp]
\centering
\caption{Evaluated files in the verification process, adapted from \cite{enase25}}
\label{tab:evaluated_files}
\resizebox{9cm}{!}{%
\begin{tabular}{|l|c|P{6cm}|}
\hline
\textbf{File} & \textbf{LOC} & \textbf{Challenges} \\ \hline
View A & 35 & Changed name to access data \\ 
View B & 25 & Array syntax access \\ 
View C & 57 & Array syntax access \\ 
View D & 190 & Contains JavaScript and PHP \\ 
View E & 19 & Ajax form to be updated to JQuery \\ 
View F & 118 & Four helper functions must be replaced \\ 
\hline

\end{tabular}%
}
\end{table}
}

%% file: tables/background.tex
\newcommand{\humanEvalmetrics}{

\begin{table}[h]
  \centering
  \caption{LLM-based systems performance compared to baseline GPT models with HumanEval metric with Pass@1}
  \label{tab:humaneval}
  \begin{tabular}{@{}P{2.5cm}@{}P{2.5cm}@{}P{2.5cm}@{}P{3.0cm}@{}r@{}}
    \hline
    \textbf{LLM-based system} & \textbf{GPT} & \textbf{HumanEval (\%)} & \textbf{Average/From baseline (\%)} & \textbf{Reference} \\
    \hline
     -	
      & GPT-3.5 turbo		& 57.3, 73.8, 74.4	& 68.5	& 	\cite{huang_agentcoder_2024,lei2024planning,zhong_debug_2024}  \\

      -	
      & GPT-4		& 67.2, 80.1, 87.2	& 78.2	& \cite{huang_agentcoder_2024,shinn2024reflexion,zhong_debug_2024}	  \\

       -
      & GPT-4o		& 90.2, 91.5	& 90,9	& \cite{lei2024planning,openai_hello_2024}  \\

      \hline

      AgentCoder & GPT-3.5 turbo		& 79.9	& 11.4	& 	\cite{huang_agentcoder_2024} \\

      LDB 
      & GPT-3.5 turbo		& 82.9	& 14.4 	& \cite{zhong_debug_2024}  \\

      LPW 	
      & GPT-3.5 turbo		& 89.0	&  20.5	& 	\cite{lei2024planning}  \\

       LDB
      & GPT 3.5/GPT-4		& 89.6	&11.4-21.1 	& \cite{zhong_debug_2024}  \\

      Reflexion
      & GPT-4		& 91.0	& 12.8	& \cite{shinn2024reflexion}  \\

      AgentCoder
      & GPT-4		& 96.3	& 18.1	& \cite{huang_agentcoder_2024}  \\

      LPW 
      & GPT-4o		& 98.1	& 7.3	& \cite{lei2024planning}  \\

      LDB \& Reflexion
      & GPT-4o		& 98.2	& 7.4	& \cite{zhong_debug_2024}  \\
    \hline
  \end{tabular}
\end{table}
}